\documentclass[runningheads]{llncs}
\usepackage{cite}
\usepackage{amsmath,amssymb,amsfonts}
\usepackage{algorithmic}
\usepackage[ruled,vlined]{algorithm2e}
\usepackage{graphicx}
\usepackage{textcomp}
\usepackage{xcolor}
\usepackage{footnotebackref}
\usepackage{multirow}
\usepackage{mathtools}
\usepackage{array,booktabs, makecell}
\usepackage{subcaption}
\usepackage{threeparttable}
\newcolumntype{M}[1]{>{\raggedright\arraybackslash}m{#1}}
\usepackage{xcolor}

\begin{document}
%
\title{LogDP: Combining Dependency and Proximity for Log-based Anomaly Detection}
%

\titlerunning{Combining Dependency and Proximity for Log-based Anomaly Detection}


\author{Yongzheng Xie\inst{1} \and
Hongyu Zhang\inst{2} \and
Bo Zhang\inst{3} \and
Muhammad Ali Babar\inst{1} \and
Sha Lu \inst{4}}

%
\authorrunning{Y. Xie et al.}
%
\institute{The University of Adelaide, Australia \email{\{yongzheng.xie,ali.babar\}@adelaide.edu.au} \and
The University of Newcastle, Australia
\email{hongyu.zhang@newcastle.edu.au} \and
The University of Newcastle, Australia
\email{c3288930@uon.edu.au} \and
University of South Australia, Australia
\email{sha.lu@mymail.unisa.edu.au}}
\maketitle              
\begin{abstract}
		Log analysis is 
		an important technique that engineers use for troubleshooting faults of large-scale service-oriented systems. In this study, we propose a novel semi-supervised log-based anomaly detection approach, LogDP, which utilizes the dependency relationships among log events and proximity among log sequences to detect the anomalies in massive unlabeled log data. LogDP divides log events into dependent and independent events, then learns normal patterns of dependent events using dependency and independent events using proximity. Events violating any normal pattern are identified as anomalies. By combining dependency and proximity, LogDP is able to achieve high detection accuracy. Extensive experiments have been conducted on real-world datasets, and the results show that LogDP outperforms six state-of-the-art methods.

\end{abstract}

	\section{Introduction} \label{section:introduction}
	Modern software-intensive systems, including service-oriented systems, have become increasingly large and complex. While these systems provide users with rich services, they also bring new challenges to system operation and maintenance. One of the challenges is to 
	identify faults and discover potential risks by analyzing a massive amount of log data. Logs are composed of semi-structured texts, i.e., log messages. Log analysis is one of the main techniques that engineers use for troubleshooting faults and capturing potential risks. When a fault occurs, checking system logs helps to efficiently detect and locate the fault. However, with the increase in scale and complexity, manual identification of abnormal logs from massive log data has become infeasible.

	During the past decade, many automated log analysis approaches, including supervised, semi-supervised, and unsupervised approaches, have been proposed to detect system anomalies reflected by logs\cite{lou2010mining,he2016experience,zhang2020anomaly,du2017deeplog,meng2019loganomaly}. Although supervised approaches show promising results, the scarcity of labeled anomalous log data is a daunting issue. In contrast, unsupervised and semi-supervised approaches have a significant advantage in that no labeled anomalous data are needed. However, the existing unsupervised and semi-supervised methods have low accuracy.
		
	In this paper, we propose a log anomaly detection method, LogDP, which simultaneously utilizes both dependency among log events and proximity among log sequences to detect anomalous log sequences. LogDP first discovers the normal patterns for logs, then identifies the log sequences that violate these patterns as anomalies. There are two types of normal patterns, dependency patterns (DPs) and proximity patterns (PPs). DPs are related to the events that have dependency relationships with other events, and PPs are for the events that are independent of other events. To find the DP of an event, LogDP trains a predictive model to predict this event using some other events as predictors. Here, we name the log event to be predicted as the \textit{focused event}, and the predictor events as the \textit{related events} of the focused event. To find the PP of an event, a mean prediction model is trained to use the mean value of the event as the expected value of the event. When detecting anomalies, given a log sequence, its expected values on all log events are predicted using the learned models, and the differences between the observed values and expected values are calculated, named \textit{pattern deviations}, which indicate the degree of the log sequence deviating from their corresponding normal dependency. If any pattern deviations are beyond normal ranges, i.e., the normal patterns are violated, the log sequence is flagged as an anomaly.

	In summary, our main contributions in this work are as follows: 
	\vspace{-6pt}
	\begin{itemize}
		\item We propose LogDP, a novel log-based anomaly detection method, which utilizes dependency among log events and proximity among log sequences at the same time. To our best knowledge, we are the first to introduce the dependency-based anomaly detection techniques in the field of log analysis. 
		
		\item We experimentally demonstrate the effectiveness of the proposed method on seven settings of three widely-used log datasets. The empirical experiments show that the proposed approach can outperform the state-of-the-art unsupervised and semi-supervised log-based anomaly detection methods.
	\end{itemize}

	\section{The LogDP Method} \label{section:methodology}	
    In this section, we first explain log preprocessing, and then present the LogDP method. The LogDP method consists of two phases, the training phase and the test phase. In the training phase, for each log event, LogDP trains an expected value prediction model and produces the corresponding threshold. In the test phase, the trained prediction modes and thresholds are used to determine if a log sequence is an anomaly or not.

    \subsection{Log Preprocessing} \label{section:parse}

\begin{figure}[tb] 
		\centering
			\begin{subfigure}[t]{0.58\textwidth}
			\centering
			\includegraphics[width=0.95\textwidth]{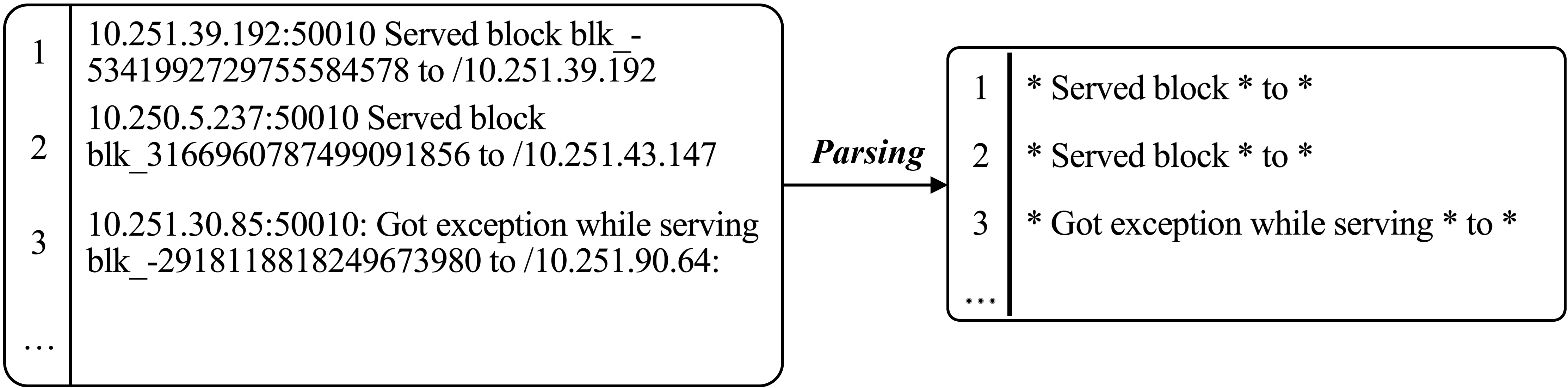}
			\caption{A snippet of log parsing.}
			\label{fig:LogsSnippet}
		\end{subfigure}
		\begin{subfigure}[t]{0.38\textwidth}
			\centering
			\includegraphics[width=0.95\textwidth]{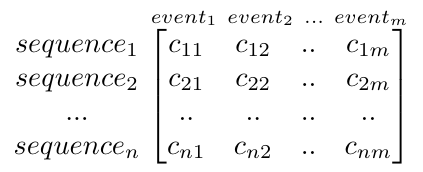}
			\caption{An ECM.}
			\label{fig:ECM}
		\end{subfigure}
		\caption{Log preprocessing.}
		\label{fig:preprocessing}
		\vspace{-12pt}
	\end{figure}

	Logs are usually semi-structured texts, which are used to record the status of systems. Each log message consists of a constant part (log event) and a variable part (log parameter). Log parsers~\cite{he2017drain, du2016spell, dai2020logram} can parse log messages into log events, which are the templates of the log messages. Figure \ref{fig:LogsSnippet} shows a snippet of raw logs and the results after they are parsed.
	
	Log messages can be grouped into log sequences (i.e., series of log events that record specific execution flows) according to sessions or time windows. Session-based log partition often utilizes certain log identifiers to generate log sequences.
	When using time windows to partition logs, two types of strategies are usually used, i.e., fixed window and sliding window. Fixed window strategy uses a predefined window size, e.g., 1 hour, to produce log sequences, while sliding windows strategy generates log sequences using overlapping between two consecutive fixed windows. For each log sequence, the occurrences of the events are counted, resulting in an Event Count Matrix (ECM). For example, an ECM is shown in Figure \ref{fig:ECM}, where $c_{ij}$ indicates the number of occurrences of $event_j$ in $sequence_i$, namely $instance_{ij}$. 
	
	The notation used in this paper is as follows. We use a boldfaced upper case letter, e.g. $\textbf{X}$ to denote a matrix; a boldfaced lower case letter, e.g. $ \textbf{e} $, for a vector; a lower case letter, e.g. $c$, for a scalar. We have reserved $\textbf{X} \in \mathbb{R}^{n \times m}$ for an ECM with $ n $ log sequences and $ m $ log events. $ \textbf{E} = \{ E_1, \cdots, E_m \} $ represents the set of log events of $ \textbf{X} $ and $ E $ is a log event, i.e., $ E \in \textbf{E} $. A log sequence is denoted as $ \textbf{c} =\{c_1, \cdots, c_m\} $, where $ c $ is a log instance, i.e., the occurrences count of an event in $ \textbf{c} $.  The log instance of event $ E_j $ in sequence $ \textbf{c}_i $ is represented as $ c_{ij} $ .

	\subsection{The Training Phase of LogDP}
\begin{figure}[tb] 
		\centering
		\includegraphics[width=1\linewidth]{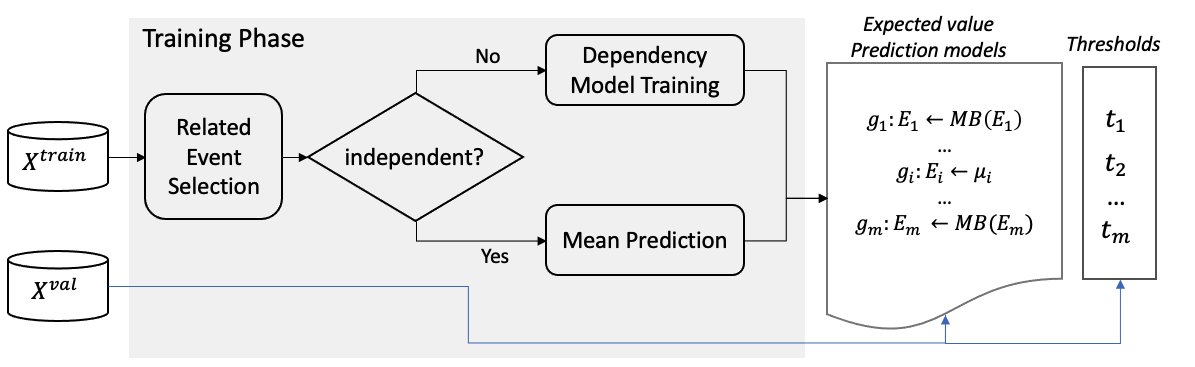}
		\caption{The workflow of the training phase of the LogDP method}
		\label{fig:training}
		\vspace{-15pt}
	\end{figure}

	The workflow of the training phase of the LogDP method is presented in Figure \ref{fig:training}. The inputs of the training phase are a training set $ \textbf{X}^{train} $ and a validation set $ \textbf{X}^{val} $, both of which only contain normal log sequences. $ \textbf{X}^{train} $ is used to train expected value prediction models, and $ \textbf{X}^{val} $ is used to obtain the thresholds. The training phase is composed of two steps, related event selection and prediction model training. In the related event selection step, for each event, named focused event, its related event is selected to be used as predictors to predict the focused event. In the prediction model training step, two different prediction models are trained according to if Markov blanket (MB) is found for the focused event. If the focused event is not independent, i.e., it has MB, a Multi-Layer Perceptron (MLP) regressor is trained to embody the dependency relationship between the focused event and its MB. If the focused event is independent, i.e., it has not MB found, a mean prediction model is trained. That is, DPs are learned for dependent events using the dependency-based technique, and PPs are for independent events using the proximity-based technique. After training the expected value prediction models, $ \textbf{X}^{val} $ is input to obtain the corresponding thresholds. The outputs of the training phase include a set of prediction models and their corresponding thresholds.
	
	\subsubsection{Related Event Selection} \label{subsection:event_selection}
	In this step, we aim to identify the related events for a focused event, which are later used as predictors in a predictive model to predict the value of a focused (independent) event. 

	We follow \cite{lu2020lopad} to adopt a causal feature selection technique, MBs, in the step to achieve a good prediction accuracy and efficiency. MBs are defined in the context of a Bayesian Network (BN) \cite{pearl2000models}. A BN is a type of probabilistic graphical model used to represent and infer the dependency among variables. In the context of log analysis, variables correspond to log events. A BN can be denoted as a pair of $(G, P)$, where $G$ is a Directed Acyclic Graph (DAG) showing the structure of the BN, and $P$ is the joint probability of the nodes in $G$. Specifically, $G=(\textbf{E},\textbf{A})$, where $\textbf{E}$ is the set of nodes representing the random variables in the domain under consideration, and $\textbf{A} \subseteq \textbf{E} \times \textbf{E}$ is the set of arcs representing the dependency among the nodes. $E_1 \in \textbf{E}$ is known as a parent of $E_2 \in \textbf{E}$ (or $E_2$ is a child of $E_1$) if there exists an arc $E_1 \rightarrow E_2$. For any variable $E \in \textbf{E}$ in a BN, its MB contains all the children, parents, and spouses (other parents of the children) of $E$, denoted as $MB(E)$. Given $MB(E)$, $E$ is conditionally independent of all other variables in $\textbf{E}$, i.e., 
	\begin{equation} \label{equ:ch3_MB}
		P(E|MB(E))= P(E|MB(E),\textbf{S}) 
	\end{equation}
	where $\textbf{S} = \textbf{E} \setminus (\{E\} \cup MB(E))$.
	
	According to Equation \ref{equ:ch3_MB}, $MB(E)$ represents the information needed to estimate the probability of $E$ by making $E$ irrelevant to the remaining variables, which makes $MB(E)$ is the minimal set of relevant variables to obtain the complete dependency of $E$. The study in \cite{lu2020lopad} has shown that using MBs as related variables could achieve better performance than other choices of related events.

	\subsubsection{Dependency Model Training} \label{subsection:model_training}
	The goal of the step is to train expected value prediction models. As shown in Figure \ref{fig:training}, after learning MBs in the first step, events are categorized into two groups, independent events, i.e., events have no MB, and dependent events, i.e., events have MB. For an independent event, the expected value is predicted as the mean of the instances of the event in the training set. For a dependent event, an MLP regressor is trained to predict the expected value of $ E $ using $ MB(E) $ as predictors. Theoretically, any regression model could be used for the step, and several regression models, such as regression trees, linear regression and SVM regressors, have been adopted in exiting dependency-base anomaly detection techniques. We chose MLP as the dependency model because it could deal with more complex data distribution and shows better performance than other regression models in our experiments.
	
	In LogDP, we consider both dependent and independent log events in anomaly detection because it is common that some anomalous messages are printed to system logs only when anomalies occur. These anomalous log messages usually have no dependency on other log events. If this case is not included in the anomaly detection, a lot of anomalies could be missed. As these anomalous events only occur when anomalies happen, they are unlikely presented in normal log sequences, which is the reason that LogDP detects them by examining the deviation from the mean of values of normal sequences.   
	
	To obtain the threshold, a validation set $ \textbf{X}^{val} $ with normal log sequences is input into the learned expected value prediction models to get the expected value of the validation set, i.e., $ \hat{\textbf{X}}^{val} $. The deviation matrix of $ \textbf{X}^{val} $ are calculated as $\textbf{D} = |\textbf{X}^{val} - \hat{\textbf{X}}^{val} | $. Then, for each event, its threshold is calculated as the maximum value of the deviations of the event, i.e., $ t_i = maximum(\textbf{D}_{*i}) $, where $ \textbf{D}_{*i} $ is the $ j $-th column of $ \textbf{D} $.
	
	\subsection{The Test Phase of LogDP}
	The goal of the test phase is to use the learned models and thresholds to detect anomalies. Given a log sequence $ \textbf{c} =\{c_i, \cdots, c_m\}$, the expected value of each instance $ c_i \in \textbf{c}$ is predicted by corresponding prediction model. Then, the deviation is calculated as $\delta =  |c_i - \hat{c}_i| $. If $ \delta > t_i$, then $\textbf{c}$ is flagged as an anomaly. $ \textbf{c} $ is considered to be normal only if it follows all the normal patterns.

	\section{Evaluation} \label{section:evaluation}

	\subsubsection{Datasets}
	Three public log datasets, HDFS, BGL and Spirit, are used in our experiments, which are available from \cite{logpai}. From the three datasets, we generate seven datasets using different log grouping strategies. The HDFS is generated using session, and BGL and Spirit are generated using 1-hour logs, 100 logs, and 20 logs windows. The names of the datasets of BGL and Spirit are denoted as Dataset-Window, e.g., BGL-100logs as shown in Table \ref{tab:datasets}.

	For LogDP, the first 2/3 sequences of the training set are used for training, and the remaining 1/3 sequences are used as a validation set.
	
	\begin{table}[tbh]
		\renewcommand{\arraystretch}{0.8}
		\centering
		\caption{Overview of datasets used in the experiments.}
		\resizebox{0.9\linewidth}{!}{%
		\begin{threeparttable}
		\begin{tabular}{ lcl cccccr}
			\toprule
			\multirow{2}{*}{Datasets} &  \multirow{2}{*}{\#Evt}  & \multirow{2}{*}{Window} & \multicolumn{3}{c}{Training Set} & \multicolumn{3}{c}{Test Set}\\
			\cmidrule{4-9} 
			& & &\#Seq  &\#Anom.  & \%Anom.  &\#Seq  &\#Anom.  &\%Anom.  \\
			\toprule
			HDFS& 29 & session&287,530 &8,419 &2.93\% & 287,531& 8,419& 2.93\% \\
			\midrule 
			& & 1 hour&3,673 &495 &13.48\% & 1,481& 170& 11.48\% \\
			\cmidrule{3-9}  
			BGL& 980 & 100 logs&37,707 &4,009 &10.63\% & 9,426& 816& 8.66\% \\
			\cmidrule{3-9}
			& & 20 logs&188,539 &17,252 &9.15\% & 47,134& 3,005& 6.38\%\\
			\midrule
			& & 1 hour&1,751 &1,213 &69.27\% & 585& 225& 38.46\%\\
			\cmidrule{3-9}
			Spirit& 1,229& 100 logs&79,999 &20,598 &25.75\% & 19,999& 429& 2.15\% \\
			\cmidrule{3-9}
			& & 20 logs&399,999 &82,002 &20.50\% & 99,999& 498& 0.50\%\\
			\bottomrule
		\end{tabular}
		\begin{tablenotes} 
		\item[1] \#Evt: number of events; \#Seq: number of sequences; \#Anom.: number of anomalies;  \%Anom.: percentage of anomalies.
		\end{tablenotes}
		\end{threeparttable} 
		}
		\label{tab:datasets}
		\vspace{-10pt}
	\end{table}
	
	\subsubsection{Benchmark Methods}
	Six state-of-the-art log-based anomaly detection methods are selected as the benchmark methods, including three proximity-based methods, PCA\cite{xu2009detecting}, OneClassSVM\cite{scholkopf2001estimating} (OCSVM), LogCluster\cite{lin2016log}; a sequential-based methods, DeepLog\cite{du2017deeplog}; and two invariant relation-based methods,  Invariant Mining\cite{lou2010mining} (IM) and ADR \cite{zhang2020anomaly}. The description of the benchmark methods can be found in Section \ref{section:related_work}.

	\subsubsection{Experimental Results}

	The experimental results (in precision, recall and F1) of LogDP and benchmark methods are presented in Table \ref{tab:results}. The best results are in boldface. Overall, LogDP produces superior results comparing to benchmark methods. Out of 7 datasets, LogDP achieves all the best results in F1; five best results in precision; two best results in recall. 
	
	\begin{table*}[tb]
        \renewcommand{\arraystretch}{0.8}
    	\centering
    	\caption{Experimental results of LogDP and benchmark methods.}
    	\label{tab:results}%
    	\resizebox{1\columnwidth}{!}{%
    		\begin{tabular}{lc ccccccr}
    			\toprule
    			\textbf{Dataset} & \textbf{Metrics} & \textbf{LogDP} & \textbf{PCA}   & \textbf{OCSVM} & \textbf{LogCluster} & \textbf{DeepLog} & \textbf{IM} & \textbf{ADR}\\
    			\toprule
    			\multirow{3}[0]{*}{HDFS-session } & F1    & \textbf{0.987} & 0.790 & 0.068 & 0.800 & 0.945 & 0.943 & 0.974 \\
    			& Precision & 0.979 & \textbf{0.980} & 0.035 & 0.870 & 0.958 & 0.893 & 0.951 \\
    			& Recall & 0.995 & 0.670 & 0.940 & 0.740 & 0.933 & \textbf{1.000} & \textbf{1.000} \\
    			\cmidrule{2-9}
    			\multirow{3}[0]{*}{	BGL-1hour } & F1    & \textbf{0.789} & 0.170 & 0.393 & 0.147 & 0.596 & 0.490 & 0.547 \\
    			& Precision & \textbf{0.935} & 0.352 & 0.383 & 0.009 & 0.474 & 0.343 & 0.377 \\
    			& Recall & 0.682 & 0.112 & 0.403 & 0.394 & 0.802 & 0.859 & \textbf{1.000}  \\
    			\cmidrule{2-9}
    			\multirow{3}[0]{*}{	BGL-100logs } & F1    & \textbf{0.539} & 0.130 & 0.132 & 0.243 & 0.378 & 0.387 & 0.250  \\
    			& Precision & \textbf{0.858} & 0.440 & 0.075 & 0.147 & 0.321 & 0.324 & 0.143  \\
    			& Recall & 0.393 & 0.076 & 0.556 & 0.705 & 0.461 & 0.482 & \textbf{0.987} \\
    			\cmidrule{2-9}
    			\multirow{3}[0]{*}{	BGL-20logs } & F1    & \textbf{0.460} & 0.237 & 0.168 & 0.226 & 0.224 & 0.203 & 0.204  \\
    			& Precision & \textbf{0.985} & 0.447 & 0.094 & 0.129 & 0.126 & 0.163 & 0.114 \\
    			& Recall & 0.300 & 0.162 & 0.744 & 0.884 & 0.981 & 0.269 & \textbf{0.988}  \\
    			\cmidrule{2-9}
    			\multirow{3}[0]{*}{	Spirit-1hour } & F1    & \textbf{0.821} & 0.187 & 0.601 & 0.367 & 0.582 & 0.387 & 0.792 \\
    			& Precision & \textbf{0.697} & 0.312 & 0.742 & 0.324 & 0.412 & 0.678 & 0.656 \\
    			& Recall & \textbf{1.000} & 0.133 & 0.505 & 0.422 & 0.991 & 0.271 & \textbf{1.000}  \\
    			\cmidrule{2-9}
    			\multirow{3}[0]{*}{	Spirit-100logs } & F1  & \textbf{0.575} & 0.111 & 0.003 & 0.110 & 0.153 & 0.107 & 0.445 \\
    			& Precision & \textbf{0.405} & 0.094 & 0.002 & 0.152 & 0.087 & 0.057 & 0.287 \\
    			& Recall & 0.993 & 0.135 & 0.023 & 0.086 & 0.643 & 0.993 & \textbf{0.994}  \\
    			\cmidrule{2-9}
    			\multirow{3}[0]{*}{	Spirit-20logs } & F1 & \textbf{0.905} & 0.095 & 0.009 & 0.173 & 0.135 & 0.032 & 0.558  \\
    			& Precision &\textbf{0.835} & 0.051 & 0.005 & 0.150 & 0.191 & 0.016 & 0.387 \\
    			& Recall & 0.988 & 0.639 & 0.057 & 0.205 & 0.104 & 0.974 & \textbf{0.999}  \\
    			\bottomrule
    		\end{tabular}%
    	}
    	\vspace{-15pt}
     \end{table*}%
     
     As for different strategies of log partitioning, i.e., session (for HDFS) or time window (for BGL and Spirit), LogDP performs well with both strategies. In contrast, as IM, ADR and DeepLog are designed to be more suitable for session-based log partitioning, they yield good results on the HDFS dataset but relatively poor results on other datasets. Compared to the benchmark methods based on proximity-based anomaly detection techniques, i.e., PCA, OCSVM and LogCluster, LogDP produces significantly better results on all datasets except for the precision of PCA on the HDFS dataset. In summary, the experiments have shown the superior performance of LogDP on different datasets with different log partition strategies.

	\section{Related Work} 	\label{section:related_work}
	Log-based anomaly detection has been intensively studied in recent decades. In terms of the techniques used for anomaly detection, the existing approach can be roughly categorized into proximity-based, sequential-based, and relation-based approaches.	Proximity-based methods, such as PCA (Principal Component Analysis)~\cite{xu2009detecting} and LogCluster~\cite{lin2016log}, cast a log event sequence, as a point in a feature space and utilize distances or density metrics to evaluate the proximity of the log sequence with others. The sequences far from the others are flagged as anomalies. Sequential-based methods, such as DeepLog~\cite{du2017deeplog} and LogAnomaly~\cite{meng2019loganomaly}, use sequences of the log events to train models and try to predict future events. The log sequences that do not comply with the predicted sequential patterns are identified as anomalies. Relation-based methods such as Invariants Mining~\cite{lou2010mining} and ADR~\cite{zhang2020anomaly}, try to find meaningful relations among the log events and use the relations to detect anomalies. 
	As a relation-based method, LogDP is more flexible than the existing ones. Existing relation-based methods \cite{lou2010mining, zhang2020anomaly} are based on the invariant relationships among log events. Invariant relations refer to the linear relationships among log events that are related to the program workflows. However, there are two limitations in the existing invariant relation-based methods: (1) the mined relations are sensitive to data noise; (2) the mined relations are restricted to linear relations among the events. In contrast, LogDP utilizes the probabilistic relationships among log events, which makes it less sensitive to data noise. LogDP also adopts MLP regressors as dependency models, which can deal with both linear and non-linear relationships.

	\section{Conclusion} \label{section:conclusion}
	We have proposed a log-based anomaly detection method, LogDP, which utilizes the deviations from normal patterns to effectively detect anomalous log sequences. LogDP divides log events into two types, dependent events and independent events. For dependent events, the normal patterns are learned from the probabilistic relationship among an event and its MB, i.e., the dependency among events. For independent events, the normal patterns are obtained from the mean prediction models, i.e., the proximity among sequences. The log sequences that violate any normal pattern are identified as anomalies. 
	Our experimental results show that LogDP outperforms the state-of-the-art benchmark methods. 
 	Our source code and experimental data are available at: \url{https://github.com/ilwoof/LogDP}.

 	\section{Acknowledgments} \label{section:acknowledgment}
    This research was supported by an Australian Government Research Training Program (RTP) Scholarship, and by the Australian Research Council's Discovery Projects funding scheme (project DP200102940). The work was also supported with super-computing resources provided by the Phoenix High Powered Computing (HPC) service at the University of Adelaide.

	\bibliographystyle{unsrt}
	\bibliography{reference}

\begin{thebibliography}{10}

\bibitem{lou2010mining}
J.~Lou, Q.~Fu, S.~Yang, Y.~Xu, and J.~Li.
\newblock Mining invariants from console logs for system problem detection.
\newblock In {\em USENIX Annual Technical Conference}, 2010.

\bibitem{he2016experience}
S.~He, J.~Zhu, P.~He, and M.~Lyu.
\newblock Experience report: System log analysis for anomaly detection.
\newblock In {\em ISSRE}, pages 207--218. IEEE, 2016.

\bibitem{zhang2020anomaly}
Bo~Zhang, Hongyu Zhang, Pablo Moscato, and Aozhong Zhang.
\newblock Anomaly detection via mining numerical workflow relations from logs.
\newblock In {\em SRDS}. IEEE, 2020.

\bibitem{du2017deeplog}
M.~Du, F.~Li, G.~Zheng, and V.~Srikumar.
\newblock Deeplog: Anomaly detection and diagnosis from system logs through
  deep learning.
\newblock In {\em Proceedings of the 2017 ACM SIGSAC Conference on Computer and
  Communications Security}, 2017.

\bibitem{meng2019loganomaly}
W.~Meng, Y.~Liu, Y.~Zhu, S.~Zhang, D.~Pei, Y.~Liu, Y.~Chen, R.~Zhang, S.~Tao,
  P.~Sun, et~al.
\newblock Loganomaly: Unsupervised detection of sequential and quantitative
  anomalies in unstructured logs.
\newblock In {\em IJCAI}, volume~19, pages 4739--4745, 2019.

\bibitem{he2017drain}
P.~He, J.~Zhu, Z.~Zheng, and M.~Lyu.
\newblock Drain: An online log parsing approach with fixed depth tree.
\newblock In {\em ICWS}. IEEE, 2017.

\bibitem{du2016spell}
Min Du and Feifei Li.
\newblock Spell: Streaming parsing of system event logs.
\newblock In {\em IEEE ICDM}, pages 859--864. IEEE, 2016.

\bibitem{dai2020logram}
H.~Dai, H.~Li, C.~Chen, W.~Shang, and T.~Chen.
\newblock Logram: Efficient log parsing using n-gram dictionaries.
\newblock {\em IEEE Transactions on Software Engineering}, 2020.

\bibitem{lu2020lopad}
S.~Lu, L.~Liu, J.~Li, T.~D. Le, and J.~Liu.
\newblock Lopad: A local prediction approach to anomaly detection.
\newblock {\em Advances in Knowledge Discovery and Data Mining}, 2020.

\bibitem{pearl2000models}
Judea Pearl.
\newblock {\em Causality: models, reasoning and inference}.
\newblock Springer, 2000.

\bibitem{logpai}
He~S., Zhu J., He~P., and R.~Lyu M.
\newblock Loghub: A large collection of system log datasets towards automated
  log analytics.
\newblock {\em arXiv e-prints}, 2020.

\bibitem{xu2009detecting}
W.~Xu, L.~Huang, A.~Fox, D.~Patterson, and M.~I Jordan.
\newblock Detecting large-scale system problems by mining console logs.
\newblock In {\em Proceedings of the ACM SIGOPS 22nd symposium on Operating
  systems principles}, pages 117--132, 2009.

\bibitem{scholkopf2001estimating}
B.~Sch{\"o}lkopf, J.~C Platt, John S-T., A.~J Smola, and R.~C Williamson.
\newblock Estimating the support of a high-dimensional distribution.
\newblock {\em Neural computation}, 2001.

\bibitem{lin2016log}
Q.~Lin, H.~Zhang, J.~Lou, Yu~Zhang, and X.~Chen.
\newblock Log clustering based problem identification for online service
  systems.
\newblock In {\em ICSE-C}. IEEE, 2016.

\end{thebibliography}

\end{document}